\documentclass[%
	    aps,
	    prr,
	    10pt,
	    superscriptaddress,
	    amsfonts,
	    amsmath,
	    amssymb,
        twocolumn,
        longbibliography
	]{revtex4-1}

\usepackage{braket}
\usepackage{hyperref}
\usepackage{xfrac}
\usepackage{xcolor}

\hypersetup{%
    pdftitle={Exploring Disordered Quantum Spin Models with a Multi-Layer Multi-Configurational Approach},
    pdfauthor={Fabian K\"{o}hler},
}

\DeclareMathOperator{\Tr}{Tr}

\begin{document}

\title{Exploring Disordered Quantum Spin Models with a Multi-Layer Multi-Configurational Approach}

\author{Fabian K\"{o}hler}
\email{fkoehler@physnet.uni-hamburg.de}
\affiliation{Center for Optical Quantum Technologies, Department of Physics, University of Hamburg, Luruper Chaussee 149, 22761 Hamburg, Germany}
\author{Rick Mukherjee}
\email{rick.mukherjee@physnet.uni-hamburg.de}
\affiliation{Center for Optical Quantum Technologies, Department of Physics, University of Hamburg, Luruper Chaussee 149, 22761 Hamburg, Germany}
\author{Peter Schmelcher}
\affiliation{Center for Optical Quantum Technologies, Department of Physics, University of Hamburg, Luruper Chaussee 149, 22761 Hamburg, Germany}
\affiliation{The Hamburg Centre for Ultrafast Imaging, University of Hamburg, Luruper Chaussee 149, 22761 Hamburg, Germany}

\begin{abstract}
    Numerical simulations of quantum spin models are crucial for a profound understanding of many-body phenomena in a variety of research areas in physics.
    An outstanding problem is the availability of methods to tackle systems that violate area laws of entanglement entropy.
    Such scenarios cover a wide range of compelling physical situations including disordered quantum spin systems among others.
    In this work, we employ a numerical technique referred to as multilayer multiconfiguration time-dependent Hartree (ML-MCTDH) to evaluate the ground state of several disordered spin models.
    ML-MCTDH has previously been used to study problems of high-dimensional quantum dynamics in molecular and ultracold physics but is here applied to study spin systems.
    We exploit the inherent flexibility of the method to present results in one and two spatial dimensions and treat challenging setups that incorporate long-range interactions as well as disorder.
    Our results suggest that the hierarchical multi-layering inherent to ML-MCTDH allows to tackle a wide range of quantum many-body problems such as spin dynamics of varying dimensionality.
\end{abstract}

\maketitle

\section{Introduction}\label{sec:introduction}

A quantum many-body system satisfies the area law of entanglement if the amount of entanglement between a subsystem and the remainder of the system is proportional to the area of the boundary~\cite{Hastings_2007_JSM2007_P08024_AreaLawOneDimensionalQuantum}.
Systems that obey the area law typically have constraints such as locality in interaction and underlying symmetries that force their eigenstates to reside on certain submanifolds of the Hilbert space, rendering their numerical simulation efficient.
Consequently, several numerical methods that rely on truncating the Hilbert space such as density matrix renormalization group method (DMRG)~\cite{White_1992_PRL69_2863_DensityMatrixFormulationQuantum,Schollwock_2005_RMP77_259_DensitymatrixRenormalizationGroup}, time evolving block decimation (TEBD)~\cite{Daley_2004_JSM2004_P04005_TimedependentDensitymatrixRenormalizationgroupUsing,Verstraete_2004_PRL93_207204_MatrixProductDensityOperators,Verstraete_2004_PRL93_207204_MatrixProductDensityOperators,White_2004_PRL93_076401_RealTimeEvolutionUsingDensity}, tensor networks~\cite{Verstraete_2008_AP57_143_MatrixProductStatesProjected,Orus_2014_AP349_117_PracticalIntroductionTensorNetworks} and other matrix product states (MPS) based methods have been very successful in simulating quantum matter for a variety of physics~\cite{White_1996_PRL77_3633_SpinGapsFrustratedHeisenberg,Kuhner_1998_PRB58_R14741_PhasesOnedimensionalBoseHubbardModel,Tezuka_2008_PRL100_110403_DensityMatrixRenormalizationGroupStudy,Stoudenmire_2012_ARCMP3_111_StudyingTwoDimensionalSystemsDensity,He_2017_PRX7_031020_SignaturesDiracConesDMRG} and chemistry problems~\cite{Legeza_2003_MP101_2019_QCDMRGStudyIonicneutralCurve,Moritz_2005_JCP123_184105_RelativisticDMRGCalculationsCurve,Liu_2013_JCTC9_4462_MultireferenceInitioDensityMatrix,Wouters_2014_JCP140_241103_CommunicationDMRGSCFStudySinglet,Wouters_2016_JCP145_054120_DMRGCASPT2StudyLongitudinalStatic,Guo_2018_JCTC14_4063_PerturbativeDensityMatrixRenormalization,Baiardi_2020_JCP152_040903_DensityMatrixRenormalizationGroup}.

However, there are quantum states that exhibit scaling of entanglement proportional to the total system size, in which case the merits of MPS based methods may be questioned.
As a matter of fact, quantum systems having strong violation of area law (entanglement grows linearly with the system size) are more common than previously expected~\cite{Vitagliano_2010_NJP12_113049_VolumelawScalingEntanglementEntropy,Pouranvari_2014_PRB89_115104_MaximallyEntangledModeMetalinsulator,Shiba_2014_JHEP2014_33_VolumeLawEntanglementEntropy,Gori_2015_PRB91_245138_ExplicitHamiltoniansInducingVolume,Roy_2018_PRB97_125116_EntanglementContourPerspectiveStrong,Roy_2019_PRA99_052342_QuantumSimulationLongrangeXY,Roy_2019_PRA100_059902E_ErratumQuantumSimulationLongRange}.
Such scenarios are typically described by disordered Hamiltonians rendering them non-translationally invariant and inducing a high level of degeneracy in their low-energy spectrum.
It is often the case that the experimental realization of many-body quantum systems are far from homogeneous, for example, crystals with dislocations or impurities~\cite{Matsudaira_1973_JPSJ35_1593_IsingFerromagnetsRandomImpurities,Oshikawa_1997_NPB495_533_BoundaryConformalFieldTheory,Frahm_1997_JPCM9_9939_OpenSpinChainImpurity}, experiments investigating quantum Hall effect~\cite{Tsui_1982_PRL48_1559_TwoDimensionalMagnetotransportExtremeQuantum,Laughlin_1983_PRL50_1395_AnomalousQuantumHallEffect,Stormer_1999_RMP71_875_NobelLectureFractionalQuantum}, glassy states of frustrated spin models~\cite{Sherrington_1975_PRL35_1792_SolvableModelSpinGlass,Edwards_1975_JPFMP5_965_TheorySpinGlasses,Binder_1986_RMP58_801_SpinGlassesExperimentalFacts} and Anderson localization~\cite{Anderson_1958_PR109_1492_AbsenceDiffusionCertainRandom,Abrahams_1979_PRL42_673_ScalingTheoryLocalizationAbsence,Roati_2008_N453_895_AndersonLocalizationNonInteractingBose}. For such systems, evaluating even the ground state can be challenging with existing methods.

In this paper, we propose an alternative numerical approach that can tackle the simulation of disordered spin systems.
The multilayer multiconfiguration time-dependent Hartree (ML-MCTDH) method~\cite{Wang_2003_JCP119_1289_MultilayerFormulationMulticonfigurationTimedependent,Manthe_2008_JCP128_164116_MultilayerMulticonfigurationalTimedependentHartree} is an extension of the MCTDH method~\cite{Meyer_1990_CPL165_73_MulticonfigurationalTimedependentHartreeApproach,Manthe_1992_JCP97_3199_WavePacketDynamicsMulticonfiguration,Beck_2000_PR324_1_MulticonfigurationTimedependentHartreeMCTDH} which was originally developed to study the multimode high-dimensional wave packet dynamics of complex molecular systems~\cite{Worth_1996_JCP105_4412_EffectModelEnvironmentS2,Huarte-Larranaga_2002_JCP116_2863_VibrationalExcitationTransitionState,Wu_2004_S306_2227_FirstPrinciplesTheoryCH4H2}.
Later extensions allow for the treatment of bosonic~\cite{Streltsov_2007_PRL99_030402_RoleExcitedStatesSplitting,Alon_2008_PRA77_033613_MulticonfigurationalTimedependentHartreeMethod,Wang_2009_JCP131_024114_NumericallyExactQuantumDynamics,Manthe_2017_JCP146_064117_MultilayerMulticonfigurationalTimedependentHartree,Weike_2020_JCP152_034101_MulticonfigurationalTimedependentHartreeApproach} and fermionic~\cite{Zanghellini_2003_LP13_1064_MCTDHFApproachMultielectronDynamics,Zanghellini_2004_JPBAMOP37_763_TestingMultiConfigurationTimeDependentHartree,Caillat_2005_PRA71_012712_CorrelatedMultielectronSystemsStrong,Sasmal_2020_JCP153_154110_NonadiabaticQuantumDynamicsPotential} ensembles as well as mixtures thereof~\cite{Cao_2013_JCP139_134103_MultilayerMulticonfigurationTimedependentHartree,Kronke_2013_NJP15_063018_NonequilibriumQuantumDynamicsUltracold,Cao_2017_JCP147_044106_UnifiedInitioApproachCorrelated}.
In an unprecedented approach, we adapt the ML-MCTDH techniques to study the ground state properties of spin models, in particular spin glass Hamiltonians which possess random couplings.
Our results show that ML-MCTDH characterizes the ground state of disordered spin systems accurately.
We demonstrate that this method can handle long-range interactions, scale to large system sizes as well as work in both one and higher spatial dimensions.
The overall flexibility of ML-MCTDH is very promising and might serve as a tool for simulating quantum many-body systems in regimes where conventional methods may falter.
Specifically it comprises the perspective of simulating the nonequilibrium quantum dynamics of many-body systems.

This paper is organized as follows.
We provide a brief introduction to ML-MCTDH in Sec.~\ref{ssec:ml-mctdh} and discuss the different spin models for which we evaluate the ground-state properties in Sec.~\ref{ssec:spin-models}.
The two prototypical disordered spin models chosen for this paper include cases of weak and strong violation of area law of entanglement entropy.
Additionally, we also include the ubiquitous transverse field Ising model with short-range and long-range interactions for comparison purposes.
Our analysis comprises ground-state characteristics such as energy, correlations, and entanglement, which are benchmarked against exact diagonalization~\cite{Lin_1990_PRB42_6561_ExactDiagonalizationQuantumspinModels} and DMRG, all of which are shown in Sec.~\ref{sec:results}.
Section~\ref{sec:conclusions-and-outlook} contains our conclusions and outlook.

\section{Theoretical Framework}\label{sec:theory}

\subsection{Multi-Layer Multi-Configuration Time-Dependent Hartree Method}\label{ssec:ml-mctdh}
To set the stage and to be self-contained, we believe it is adequate and instructive to provide a brief introduction to the ML-MCTDH method.
One of the main challenges in the numerical treatment of quantum many-body systems is the exponential growth of Hilbert space dimension with system size.
In this section, we describe how the ML-MCTDH is able to represent complex many-body wave functions with many degrees of freedom and thus deal with large system sizes.
We start by first discussing the original MCTDH method, which already contains the fundamental working principles and extend to ML-MCTDH by adding the notion of a hierarchy of multiple layers.
\begin{figure*}[t!]
    \centering
    \includegraphics[width=2\columnwidth]{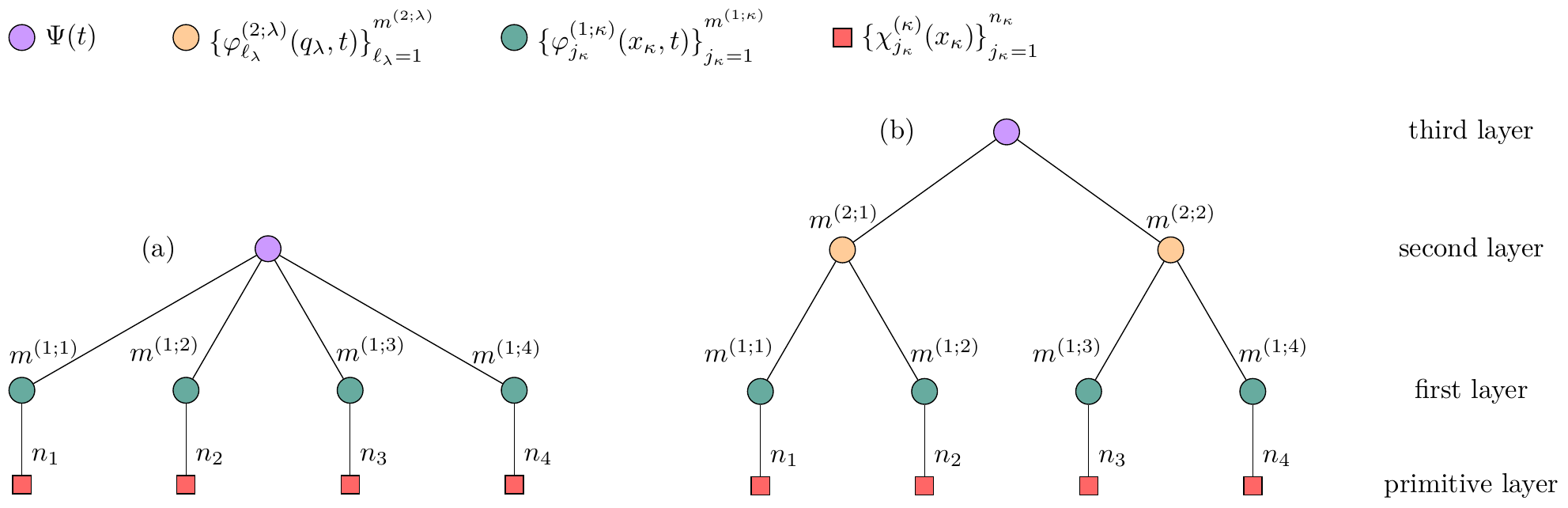}
    \caption{Diagrammatic representation of a three-layer MCTDH (a) and a four-layer ML-MCTDH (b) ansatz for the many-body wave function $\Psi(t)$ of a system with $N=4$ physical degrees of freedom.}
    \label{fig:setup}
\end{figure*}

The traditional and most straightforward approach to wave packet dynamics uses an ansatz given by a linear superposition $\ket{\Psi(t)}=\sum_J A_J(t)\ket{\Phi_J}$ of time-independent $\ket{\Phi_J}$ configurations with time-dependent coefficients $A_J(t)$.
Without loss of generality, we assume a physical scenario with $N$ degrees of freedom $x_{\kappa}$ with $\kappa=1,\ldots,N$.
Depending on the system under consideration, the degrees of freedom could, for example, be spatial degrees of freedom of particles or bosonic/fermionic occupation numbers.
A set of $n_{\kappa}$ time-independent (primitive) basis functions $\ket{\chi_{j_{\kappa}}^{(\kappa)}(x_{\kappa})}$ with $j_{\kappa}=1,\ldots,n_{\kappa}$ is employed for each degree of freedom.
The $\ket{\chi_{j_{\kappa}}^{(\kappa)}(x_{\kappa})}$ are naturally chosen to form an orthonormal basis for each degree of freedom.
The configurations $\ket{\Phi_J}$ are product states with respect to combinations of the primitive basis functions where the multi-index $J=\left(j_1,j_2,\ldots, j_{N}\right)$ runs through all possible combinations such that the full wave function ansatz is given by
\begin{equation}\label{eq:standard-ansatz-expanded}
    \ket{\Psi(t)}=\sum\limits_{j_1=1}^{n_1}\cdots\sum\limits_{j_N=1}^{n_N}A_{j_1\ldots j_N}(t)\bigotimes\limits_{\kappa=1}^{N}\ket{\chi_{j_\kappa}^{(\kappa)}(x_{\kappa})}.
\end{equation}
The time evolution of the many-body wave function $\ket{\Psi(t)}$ is governed by the Dirac-Frenkel variational principle~\cite{Dirac_1930_MPC26_376_NoteExchangePhenomenaThomas,Frenkel_1934_WaveMechanicsAdvancedGeneral},
\begin{equation}\label{eq:tdvp}
    \Braket{\delta\Psi(t)|\left(\imath\partial_t-\hat{H}\right)|\Psi(t)}=0.
\end{equation}
By inserting the wave function ansatz~\eqref{eq:standard-ansatz-expanded} in Eq.~\eqref{eq:tdvp}, one obtains the equation of motion for the expansion coefficients $A_J(t)$,
\begin{equation}
    \imath\dot{A}_J(t)=\sum\limits_L \braket{\Phi_J|\hat{H}|\Phi_L}A_L(t)
\end{equation}
which can be solved numerically using standard time integration methods.

In this traditional wave packet ansatz, the number of configurations and corresponding coefficients $\prod_{\kappa=1}^N n_{\kappa}$, scales exponentially with $N$, limiting the applicability of this approach to systems with only few degrees of freedom.
In many physical scenarios, it is often the case that using a small set of \textit{time-dependent} basis functions can provide an accurate representation of the many-body wave function thereby allowing to simulate larger systems.
Thus, in MCTDH, Eq.~\eqref{eq:standard-ansatz-expanded} is replaced with time-dependent configurations,
\begin{equation}\label{eq:mcthd-ansatz}
    \ket{\Psi(t)}=\sum\limits_{j_1=1}^{m^{(1;1)}}\cdots\sum\limits_{j_N=1}^{m^{(1;N)}}A_{j_1\ldots j_N}^{(1)}(t)\bigotimes\limits_{\kappa=1}^{N}\ket{\varphi_{j_\kappa}^{(1;\kappa)}(x_{\kappa},t)},
\end{equation}
where $\ket{\varphi_{j_\kappa}^{(1;\kappa)}(x_{\kappa},t)}$ denotes the $j_{\kappa}$th time-dependent basis function for the $\kappa$th degree of freedom and are referred to as single particle functions (SPFs).
The numbers $m^{(1;\kappa)}$ specify the number of SPFs used for the $\kappa$th degree of freedom.
The superscript $(1)$ or $(1;\kappa)$ for the SPFs, coefficients and SPF numbers indicate that these objects are part of the same, first layer of the wave function ansatz, a notation that will become essential for the multilayer extension below.
The SPFs in turn are represented with respect to the time-independent basis of the standard ansatz~\eqref{eq:standard-ansatz-expanded},
\begin{equation}\label{eq:mctdh-spfs}
    \ket{\varphi_{j_{\kappa}}^{(1;\kappa)}(x_{\kappa},t)}=\sum\limits_{\ell=1}^{n_\kappa}c_{j_{\kappa};\ell}^{(\kappa)}(t)\ket{\chi_{\ell}^{(\kappa)}(x_{\kappa})}.
\end{equation}
The MCTDH wave function ansatz can be understood as a three-layer approach [see Fig.~\ref{fig:setup} (a)].
The top layer corresponds to the total many-body wave function expanded with respect to the SPFs using time-dependent coefficients.
The middle layer refers to the time-dependent SPFs expanded with respect to the time-independent primitive basis functions while the lowest layer contains the primitive basis functions themselves.
The time-dependent variational principle~\eqref{eq:tdvp} yields equations of motion for both the coefficients $A_{j_1\ldots j_N}^{(1)}(t)$ and the SPFs $\ket{\varphi_{j_{\kappa}}^{(1;\kappa)}(x_{\kappa},t)}$, which we omit here for brevity but more details can be found in Ref.~\cite{Beck_2000_PR324_1_MulticonfigurationTimedependentHartreeMCTDH}.
In order to ensure convergence, a sufficient number of SPFs has to be employed such that they span a Hilbert space of adequate size in order to capture the underlying physics correctly.
As a matter of fact, it is often the case that the MCTDH wave function ansatz~\eqref{eq:mcthd-ansatz} contains a much smaller number of configurations compared to the wave packet ansatz~\eqref{eq:standard-ansatz-expanded}, i.e., $\prod_{\kappa=1}^{N}m_{\kappa}\ll\prod_{\kappa=1}^N n_{\kappa}$, leading to a significant reduction of the computational effort.
MCTDH was successfully used to study molecular problems with 12--14 degrees of freedom~\cite{Worth_1996_JCP105_4412_EffectModelEnvironmentS2,Huarte-Larranaga_2002_JCP116_2863_VibrationalExcitationTransitionState,Wu_2004_S306_2227_FirstPrinciplesTheoryCH4H2} and later extended to 15--24 degrees of freedom~\cite{Raab_2000_CPL319_674_DiracFrenkelMclachlanVariational,Cattarius_2001_JCP115_2088_AllModeDynamicsConical,Vendrell_2007_JCP127_184302_Fulldimensional15dimensionalQuantumdynamicalSimulation,Vendrell_2007_JCP127_184303_FullDimensional15dimensionalQuantumdynamical} and even 100 degrees of freedom for system-bath problems~\cite{Wang_2000_JCP113_9948_BasisSetApproachQuantum,Wang_2001_JCP115_2979_SystematicConvergenceDynamicalHybrid,Nest_2003_JCP119_24_DissipativeQuantumDynamicsAnharmonic} using mode combination~\cite{Ehara_1996_JCP105_8865_MulticonfigurationTimeDependentHartree,Worth_1998_JCP109_3518_RelaxationSystemConicalIntersection}.
However, capturing beyond-mean-field effects requires at least two SPFs for each degree of freedom such that the total number of configurations is at least $2^N$, highlighting the exponential scaling with respect to the system size.

In order to treat much larger systems, the ML-MCTDH approach was introduced, which has been highly successful in the treatment of systems with hundreds or even thousands of degrees of freedom~\cite{Wang_2003_JCP119_1289_MultilayerFormulationMulticonfigurationTimedependent,Wang_2006_JCP125_174502_CalculationReactiveFluxCorrelation,Wang_2007_JPCA111_10369_QuantumDynamicalSimulationElectronTransfer,Kondov_2007_JPCC111_11970_QuantumDynamicsPhotoinducedElectronTransfer} including the study of vibrational as well as electronic dynamical processes in molecules~\cite{Westermann_2011_JCP135_184102_PhotodissociationMethylIodideEmbedded,Schulze_2015_JPCB119_6211_ExplicitCorrelatedExcitonVibrationalDynamics} or linear rotor chains~\cite{Mainali_2021_JCP154_174106_ComparisonMultilayerMulticonfigurationTimedependent}.
The central idea of ML-MCTDH is to group the $N$ physical degrees of freedom $x_1,\ldots x_N$ into $d$ logical coordinates as shown below,
\begin{equation}
    \begin{aligned}
        q_1 & =\lbrace x_1,x_2,\ldots,x_{s_1}\rbrace                 \\
        q_2 & =\lbrace x_{s_1+1},x_{s_1+2},\ldots,x_{s_1+s_2}\rbrace \\
            & \vdots                                                 \\
        q_d & =\lbrace x_{s_1+\ldots+s_{d-1}+1},\ldots,x_{N}\rbrace.
    \end{aligned}
\end{equation}
For each logical coordinate $q_{\lambda}$ a new set of time-dependent SPFs ${\lbrace\ket{\varphi_{\ell_{\lambda}}^{(2;\lambda)}(q_{\lambda},t)}\rbrace}_{\ell_{\lambda}=1}^{m^{(2;\lambda)}}$ is introduced.
In ML-MCTDH, the many-body wave function ansatz Eq.~\eqref{eq:mcthd-ansatz} is replaced by expanding it with respect to these new, second layer SPFs
\begin{equation}
    \ket{\Psi(t)}=\sum\limits_{\ell_1=1}^{m^{(2;1)}}\cdots\sum\limits_{\ell_d=1}^{m^{(2;d)}}A_{\ell_1,\ldots,\ell_d}^{(2)}(t)\bigotimes\limits_{\lambda=1}^{d}\ket{\varphi_{\ell_{\lambda}}^{(2;\lambda)}(q_{\lambda},t)}\label{eq:mlmctdh_ansatz}.
\end{equation}
The newly introduced functions $\ket{\varphi_{\ell_{\lambda}}^{(2;\lambda)}(q_{\lambda},t)}$ are represented with respect to a subset of the original MCTDH SPFs given by Eq.~\eqref{eq:mctdh-spfs} that are associated with the logical coordinate $q_{\lambda}$, i.e.,
\begin{equation}
    \begin{aligned}
        \ket{\varphi_{\ell_{\lambda}}^{(2;\lambda)}(q_{\lambda},t)}=\sum\limits_{j_{\alpha}=1}^{m^{(1;\alpha)}} & \cdots\sum\limits_{j_{\beta}=1}^{m^{(1;\beta)}}\biggl[ A_{\ell_{\lambda};j_{\alpha},\ldots,j_{\beta}}^{(1;\lambda)}(t) \\
        \cdot                                                                                                   & \bigotimes\limits_{\kappa=\alpha}^{\beta}\ket{\varphi_{j_{\kappa}}^{(1;\kappa)}(x_{\kappa},t)}\biggr].
    \end{aligned}
\end{equation}
Here, $\alpha=\alpha(\lambda)=1+\sum_{i=1}^{\lambda-1}s_i$ and $\beta=\beta(\lambda)=\sum_{i=1}^{\lambda}s_i$ correspond to the index of the first and last physical coordinate associated with the logical coordinate $q_{\lambda}$ respectively.
The newly introduced SPFs $\ket{\varphi_{\ell_{\lambda}}^{(2;\lambda)}(q_{\lambda},t)}$ can be interpreted as a multidimensional wave function that follows an MCTDH ansatz with respect to the original MCTDH SPFs~\eqref{eq:mctdh-spfs}.
With this interpretation, ML-MCTDH can be viewed as adding another layer to the original MCTDH scheme ending up in a four-layer ansatz for the many-body wave function, which is schematically depicted in Fig.~\ref{fig:setup}(b).
In general, more middle layers can be added where each layer introduces a new set of SPFs that are constructed using an MCTDH ansatz with respect to the layer below in a recursive manner.
This allows the tree structure to be adapted and tailored specifically for the physical problem under consideration.
It should be noted that the SPFs across all layers are chosen to form orthonormal basis sets and remain orthonormal throughout the time evolution.
In summary, ML-MCTDH offers great flexibility regarding the degrees of freedom due to the choice of an appropriate primitive basis according to the physical problem under consideration.
When treating the dynamics of particles for example, FFT-based~\cite{Kosloff_1983_JCP52_35_FourierMethodSolutionTime,Kosloff_1988_JPC92_2087_TimedependentQuantummechanicalMethodsMolecular} schemes or discrete variable representations~\cite{Harris_1965_JCP43_1515_CalculationMatrixElementsOneDimensional,Dickinson_1968_JCP49_4209_CalculationMatrixElementsOneDimensional,Light_1985_JCP82_1400_GeneralizedDiscreteVariableApproximation} are commonly used to provide a primitive basis for the spatial degrees of freedom.
By using fermionic~\cite{Sasmal_2020_JCP153_154110_NonadiabaticQuantumDynamicsPotential} or bosonic~\cite{Wang_2009_JCP131_024114_NumericallyExactQuantumDynamics,Manthe_2017_JCP146_064117_MultilayerMulticonfigurationalTimedependentHartree,Weike_2020_JCP152_034101_MulticonfigurationalTimedependentHartreeApproach} occupation numbers the treatment of indistinguishable particles is possible as well.

In the present paper, we investigate spin-$\sfrac{1}{2}$ systems and consequently employ a two-dimensional primitive basis containing the spin-up and spin-down state for each degree of freedom, i.e., $n_{\kappa}=2$, $\ket{\chi_1^{(\kappa)}}=\ket{\uparrow}$ and $\ket{\chi_2^{(\kappa)}}=\ket{\downarrow}$ for all $\kappa=1,\ldots,N$.
While in general MCTDH and ML-MCTDH are tools to study the dynamics of many-body quantum systems, they also provide access to eigenstates of the underlying Hamiltonian by switching from real to imaginary time propagation. More details can be found in Appendix~\ref{app:computation_of_eigenstates}.

\subsection{Spin Models}\label{ssec:spin-models}

\begin{figure*}[t!]
    \includegraphics[width=2\columnwidth]{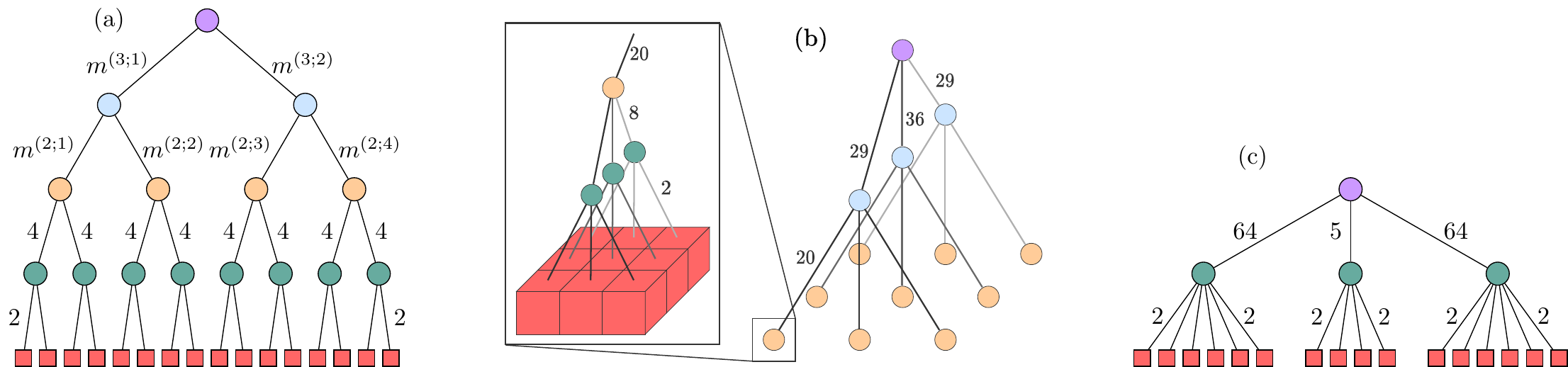}
    \caption{%
    (a) Tree structure used for the SR-TFIM ($m^{(3;i)}=6$ and $m^{(2;i)}=12$), LR-TFIM ($m^{(2;i)}=m^{(3;i)}=16$) as well as XY-SG ($m^{(2;i)}=16$ and $m^{(3;i)}=32$) of $L=16$ spins in 1D.
    (b) Tree structure used for the SR-TFIM extended to 2D on a square lattice of $9\times 9$ spins.
    (c) Tree structure used for the SDRG model of $L=16$ spins in 1D.
    }
    \label{fig:trees}
\end{figure*}
Three different quantum spin models are investigated in order to study the performance of the ML-MCTDH method.
As a starting point and for comparison purposes, it is useful to consider the transverse field Ising model (TFIM)~\cite{Pfeuty_1970_AP57_79_OnedimensionalIsingModelTransverse,Suzuki_2012_QuantumIsingPhasesTransitions} as it is one of the most fundamental and well studied models and has been realized in a variety of physical setups including trapped ions~\cite{Porras_2004_PRL92_207901_EffectiveQuantumSpinSystems,Kim_2009_PRL103_120502_EntanglementTunableSpinSpinCouplings,Britton_2012_N484_489_EngineeredTwodimensionalIsingInteractions,Islam_2013_S340_583_EmergenceFrustrationMagnetismVariableRange}, Rydberg atoms~\cite{Weimer_2010_NP6_382_RydbergQuantumSimulator,Lesanovsky_2011_PRL106_025301_ManyBodySpinInteractionsGround,Schauss_2015_S347_1455_CrystallizationIsingQuantumMagnets,Labuhn_2016_N534_667_TunableTwodimensionalArraysSingle}, and single crystals~\cite{Coldea_2010_S327_177_QuantumCriticalityIsingChain}.
The Hamiltonian of the TFIM in 1D is given by
\begin{equation}
    H_{\text{TFIM}}=
    -\sum\limits_{\substack{i,j=1 \\ i<j}}^{L}J_{ij}\sigma_i^z\sigma_{j}^z
    -h_x\sum\limits_{i=1}^{L}\sigma_i^x
    -h_z\sum\limits_{i=1}^{L}\sigma_i^z
    \label{eq:hamiltonian-tfim}
\end{equation}
where $J_{ij}$ specifies the interaction strength between the $i$th and $j$th spin while $h_x$ ($h_z$) determines the strength of a transverse (longitudinal) magnetic field.
We consider both nearest-neighbor interactions (SR-TFIM), i.e., $J_{ij}=J\delta_{i+1,j}$, and long-range interactions that decay as a power law of the distance between the spins, i.e., $J_{ij}=J{\left|i-j\right|}^{-\alpha}$ (LR-TFIM).
The parameter $J$ determines the energy scale of the system and the exponent $\alpha$ controls the range of the interactions.
We choose $J>0$ such that ferromagnetic order, i.e., the alignment of neighboring spins in the $z$ direction, is energetically favorable.
For the long-range interactions, we choose $\alpha=3$, which is accessible by trapped ions as well as Rydberg atoms.

For the remaining two models in the present paper (see below) we choose disordered systems that violate the area law of entanglement entropy.
Numerical methods like DMRG, which are based on matrix product states rely on the area law and may fall short while treating such models.
While it has been shown that a homogeneous, gapped 1D spin systems with local interactions like the SR-TFIM obey the area law~\cite{Hastings_2007_JSM2007_P08024_AreaLawOneDimensionalQuantum,Eisert_2010_RMP82_277_ColloquiumAreaLawsEntanglement}, understanding the impact of disorder on the entanglement properties of ground states remains an open and challenging question.
It is known that in such nontranslationally invariant scenarios, weak (logarithmic scaling with the system size)~\cite{Eisert_2010_RMP82_277_ColloquiumAreaLawsEntanglement,Refael_2004_PRL93_260602_EntanglementEntropyRandomQuantum,Turkeshi_2020_PRB102_014455_EntanglementEquipartitionCriticalRandom} or even stronger~\cite{Vitagliano_2010_NJP12_113049_VolumelawScalingEntanglementEntropy,Pouranvari_2014_PRB89_115104_MaximallyEntangledModeMetalinsulator,Shiba_2014_JHEP2014_33_VolumeLawEntanglementEntropy,Gori_2015_PRB91_245138_ExplicitHamiltoniansInducingVolume,Roy_2018_PRB97_125116_EntanglementContourPerspectiveStrong,Roy_2019_PRA99_052342_QuantumSimulationLongrangeXY,Roy_2019_PRA100_059902E_ErratumQuantumSimulationLongRange} violations of the area law can occur.
Our first disordered model is a XY spin glass (XYSG)~\cite{Albrecht_1982_PRL48_819_HeisenbergXYIsingSpinGlass,DeCesare_1990_PLA145_291_TwospinClusterApproachInfiniterange,DeCesare_1992_PRB45_1041_CavityfieldsApproachQuantumXY} given by the Hamiltonian
\begin{equation}
    H_{\text{XYSG}}=
    \sum\limits_{\substack{i,j=1\\i<j}}^{L}\frac{J_{ij}}{{\left|i-j\right|}^{\alpha}}
    \left(
    \sigma_i^+\sigma_j^-
    + \sigma_j^+\sigma_i^-
    \right)
    \label{eq:hamiltonian-spin-glass}
\end{equation}
with the spin flip operators $\sigma^{\pm}=\sigma^x\pm\imath\sigma^y$.
We choose $\alpha=3$ and $J_{ij}$ from a uniform distribution in $\left[-1, 1\right]$.
This spin glass model exhibits weak violation of the area law~\cite{Roy_2019_PRA99_052342_QuantumSimulationLongrangeXY,Roy_2019_PRA100_059902E_ErratumQuantumSimulationLongRange}.
The second disordered spin model we analyze is motivated by the strong disorder renormalization group (SDRG) framework~\cite{Young_1976_JPCSSP9_4419_RealspaceRenormalizationGroupCalculations,Igloi_2005_PR412_277_StrongDisorderRGApproach,Vitagliano_2010_NJP12_113049_VolumelawScalingEntanglementEntropy,Igloi_2012_PRB85_094417_EntanglementEntropyDynamicsDisordered,Vosk_2014_PRL112_217204_DynamicalQuantumPhaseTransitions} whose ground state is known to exhibit strong area-law violation.
The relevant Hamiltonian is
\begin{equation}	\label{eq:hamiltonian-sdrg}
    H_{\text{SDRG}}=
    \frac{1}{2}\sum\limits_{i=1}^{L-1} J_i
    \left(
    \sigma_i^x\sigma_{i+1}^x
    +\sigma_i^y\sigma_{i+1}^y
    \right) ,
\end{equation}
where the spin couplings are fine-tuned to be $J_i=J_0 f\left(\left|\sfrac{L}{2}-i\right|\right)$ with $f(n)=e^{-2n^2}$~\cite{Vitagliano_2010_NJP12_113049_VolumelawScalingEntanglementEntropy}.
In general, a 1D spin chain with nearest-neighbor interactions like the SR-TFIM~\eqref{eq:hamiltonian-tfim} can be solved exactly by mapping it to the free fermionic chain via the Jordan-Wigner transformation.
Models that incorporate disorder or long-range interactions like Eqs.~\eqref{eq:hamiltonian-spin-glass} and~\eqref{eq:hamiltonian-sdrg} cannot be treated this way rendering the development of powerful numerical tools like ML-MCTDH crucial.

\emph{A priori,} it is not clear which tree structure is best suited to treat a given many-body problem with the ML-MCTDH method.
In particular, different topologies can lead to vastly different simulation runtimes but yield comparable results as long as proper convergence with respect to the number of SPFs on each layer is ensured.
Finding a good tree structure is an iterative process that is guided by monitoring the occupation of the SPFs as well as the physical observables under consideration.
As a starting point, it is usually beneficial to couple degrees of freedom at the lowest layers of the tree that are strongly interacting in the underlying Hamiltonian.
The goal is to exploit the multilayering aspect of the method as much as possible in order to obtain a very compact representation of the many-body wave function and thus reduce the computational cost.
In Fig.~\ref{fig:trees} we show the various tree diagrams that are used in the present paper.
For the SR-TFIM we employ a binary tree with $\log_2(L)+1$ layers, see panel (a).
This choice is natural as it couples the neighboring spins on the lowest layers.
Since this cannot be achieved for all couplings at the same time, some of these interactions are mediated through the upper layers.
The same binary tree topology works for the LR-TFIM as well since the interaction between neighboring spins is still the strongest.
However, due to the long-range character of the interactions, more SPFs have to be used on the upper layers in order to capture long-range effects.
A binary tree structure also works well for describing the XYSG model where the design of a more optimized tree structure is prohibitive due to the random nature of the couplings.
When treating two-dimensional systems more complex tree structures are required [see panel (b)].
In the present example of a $9\times 9$ square lattice, we alternate between combining triplets of logical coordinates along the $x$ and $y$ direction.
We can treat the SDRG model accurately with the tree depicted in panel (c), which is a simple MCTDH ansatz with mode combination that does not rely on any multilayering.
This approach combines the strongly interacting central spins into one logical coordinate, which is then coupled to logical coordinates combining the outer spins.

\section{Results and Discussion}\label{sec:results}

\begin{figure}[t!]
    \centering
    \includegraphics[width=\columnwidth]{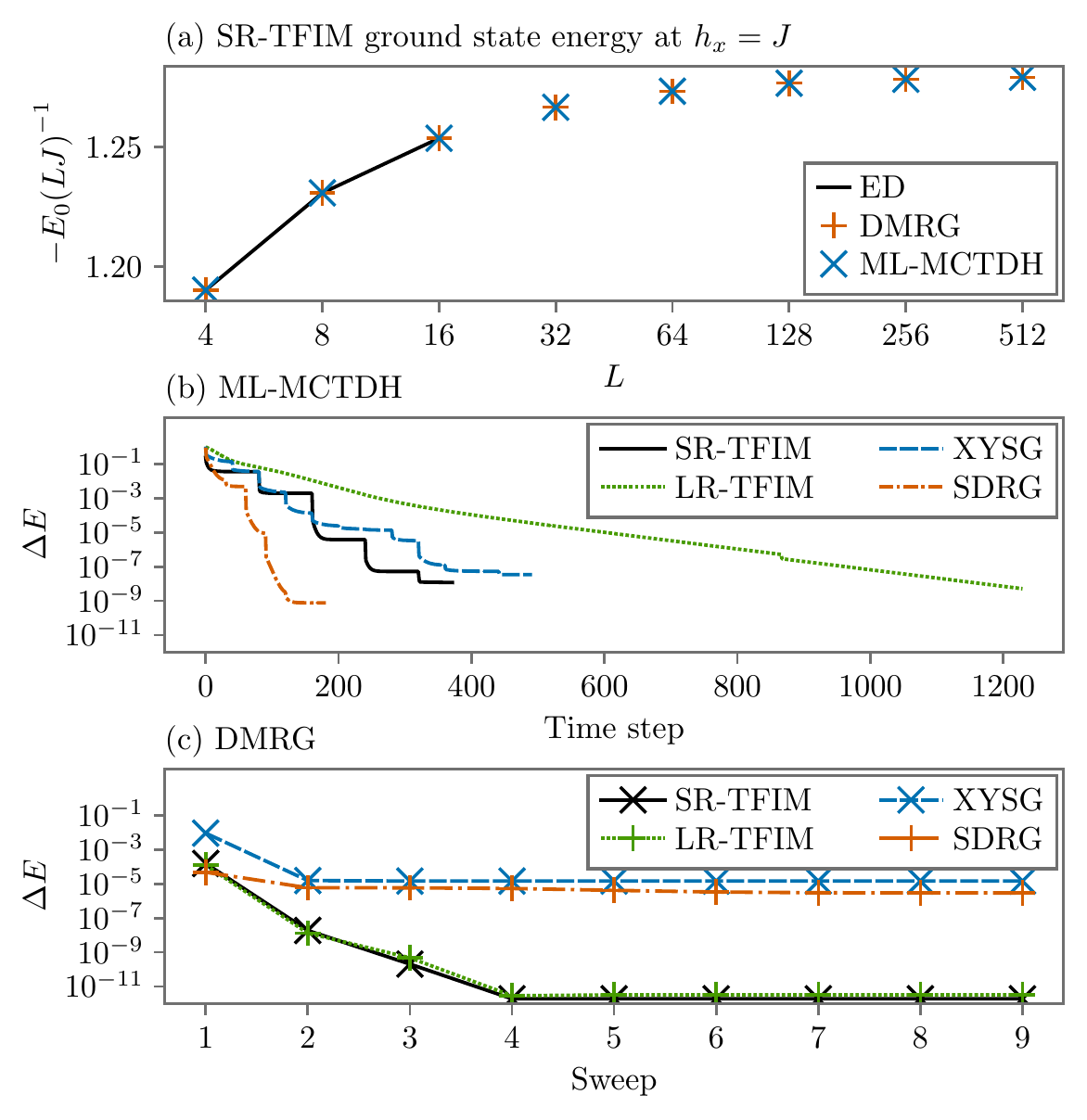}
    \caption{%
        (a) Ground-state energy per spin of the SR-TFIM in 1D for $h_x=J$ and $h_z=0.01J$ as a function of the system size $L$.
        (b) Relative error of the ML-MCTDH ground-state energy with respect to the ED ground-state energy as a function of imaginary time step for different 1D models of $L=16$ spins.
        (c) Relative error of the DMRG ground-state energy with respect to ED ground-state energy as a function of the sweep index for different 1D models of $L=16$ spins.
    }
    \label{fig:energy}
\end{figure}
We benchmark the performance of the ML-MCTDH method against exact diagonalization (ED) and DMRG by characterizing the ground state of different spin models using its energy $E_0$, correlation functions $C_{\beta\beta}(i,j)$, and entanglement entropy $S_{\mathrm{vN}}$.
The exact diagonalization implementation uses the QuSpin package~\cite{Weinberg_2017_SP2_003_QuSpinPythonPackageDynamics} in conjunction with some routines provided by quimb~\cite{Gray_2018_JOSS3_819_QuimbPythonPackageQuantum}.
The DMRG code is based on the ITensor library~\cite{Fishman_2022_SPC_004_ITensorSoftwareLibraryTensor}.

Figure~\ref{fig:energy}(a) shows the ground-state energy per spin for the SR-TFIM as a function of system size $L$ for a fixed transverse field of $h_x=J$ for which there is excellent agreement between all three methods.
Naturally ED is limited to a few spins, while DMRG and ML-MCTDH can treat much longer chains, exhibiting great scalability with respect to the system size.
However, calculating ground states for large systems can be computationally time consuming.
In order to accelerate the convergence to the ground state for these large systems, we impose a small longitudinal magnetic field $h_z=0.01J$, which lifts the twofold degeneracy of the ground state.
It should be noted that our approach works as well in the absence of a longitudinal field.
Figure \ref{fig:energy}(b) and \ref{fig:energy}(c) illustrate the convergence of the ground-state energy $E_{0;\text{M}}$ obtained by ML-MCTDH and DMRG with respect to the ground-state energy $E_{0,\text{ED}}$ computed with ED.
This is quantified by calculating the relative error $\Delta E=\left|\sfrac{E_{0;\text{M}}}{E_{0;\text{ED}}}-1\right|$ as a function of time steps for ML-MCTDH in (b) and number of sweeps for DMRG in (c).
When compared to ED, both methods achieve excellent accuracy for the SR- and LR-TFIM, but for the disordered XYSG and SDRG systems, it is clear that ML-MCTDH manages to obtain a much higher precision than DMRG.

Figure~\ref{fig:energy} (c) also illustrates that the DMRG ground-state energy converges rapidly and reaches its final value already after $2$--$4$ sweeps.
We employ a protocol consisting of nine sweeps and allow the bond dimension of the matrix product states to dynamically grow up to $1000$.
More details on this scheme can be found in Appendix~\ref{app:dmrg_protocol}.
A maximal bond dimension of $14$ for the SR-TFIM and $57$ for the LR-TFIM of $L=16$ spins is sufficient for an accurate description of the ground state across the whole range of transversal fields.
The XYSG demands a higher maximal bond dimension of $129$ due to its disordered character.
The SDRG model requires a surprisingly low final maximal bond dimension of $8$.
By forcing the DMRG algorithm to use a minimal bond dimension of at least $100$ and checking all observables under consideration, we ensured that our results for the SDRG model are indeed converged and an increase in bond dimension does not improve the results.

\begin{figure}[ht!]
    \centering
    \includegraphics[width=\columnwidth]{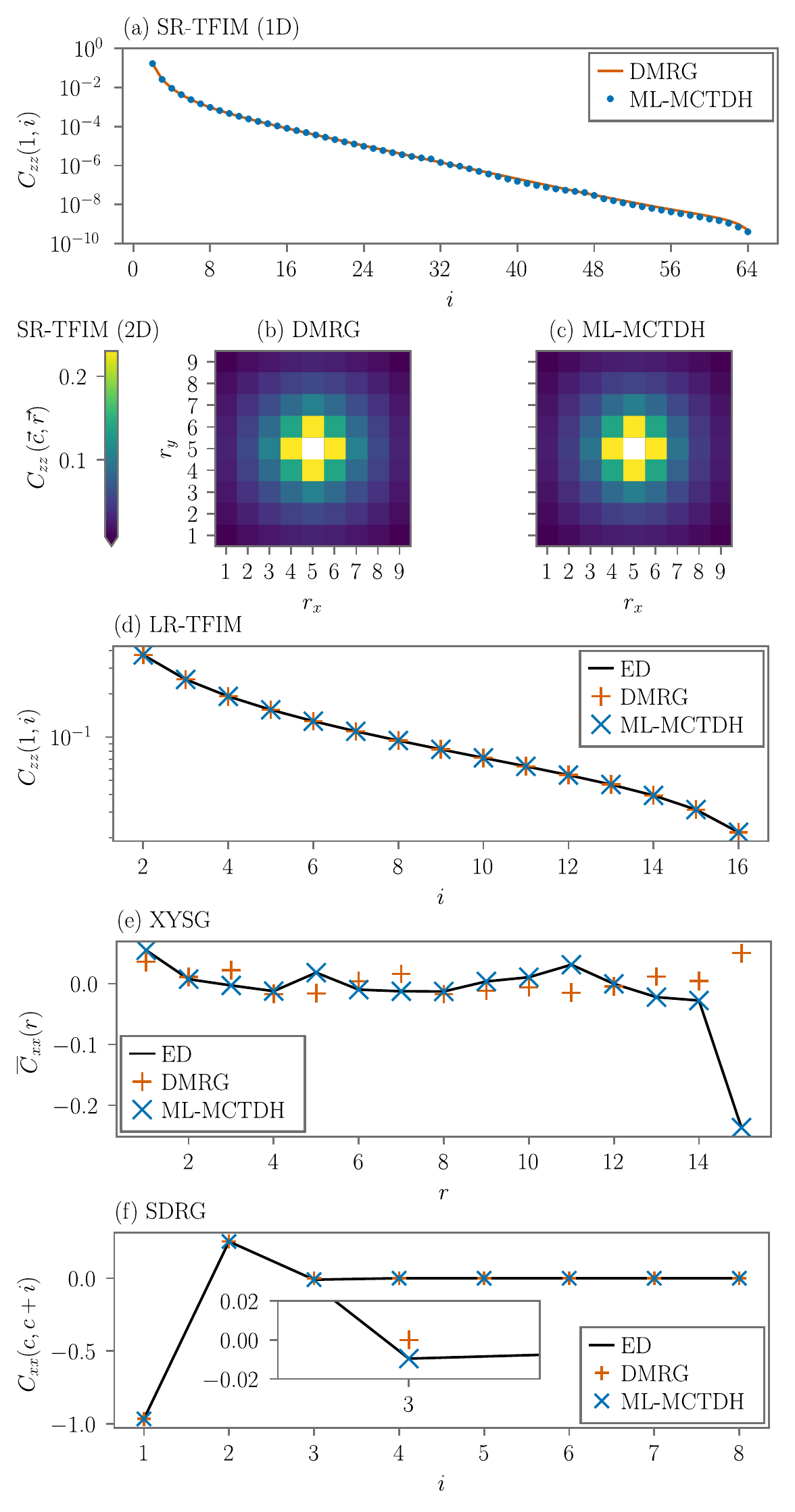}
    \caption{Connected correlation~\eqref{eq:connected-correlation-functions} functions for different models.
        (a) Correlation of the first spin with the $i$th spin in $z$ direction for the SR-TFIM of $L=128$ spins in 1D with $h_x=1.5J$ and $h_z=0.01J$.
            [(b),(c)] Correlation of the central spin with the spin at position $\vec{r}$ in $z$ direction for the SR-TFIM extended to 2D on a $9\times 9$ square lattice (2D) for $h_x=3J$ and $h_z=0.01J$.
        (d) Correlation in $z$ direction of the first spin with the $i$th spin in the LR-TFIM for $L=16$ in 1D with $h_x=J$ and $h_z=0$.
        (e) Correlation of the first spin in the XYSG for $L=16$ in 1D.
        We average over $10$ disorder realizations as well as all unique spin pairings $C_{xx}(i,j)$ with $i<j$ that correspond to a given separation distance $r=\left|i-j\right|$ and show the result as a function of $r=$.
        (f) Correlation in $x$ direction of one of the central spins $c=\sfrac{L}{2}$ with its right-hand side neighbors for the SDRG model with $L=16$ in 1D.
    }
    \label{fig:correlations}
\end{figure}
\begin{figure}[ht!]
    \centering
    \includegraphics[width=\columnwidth]{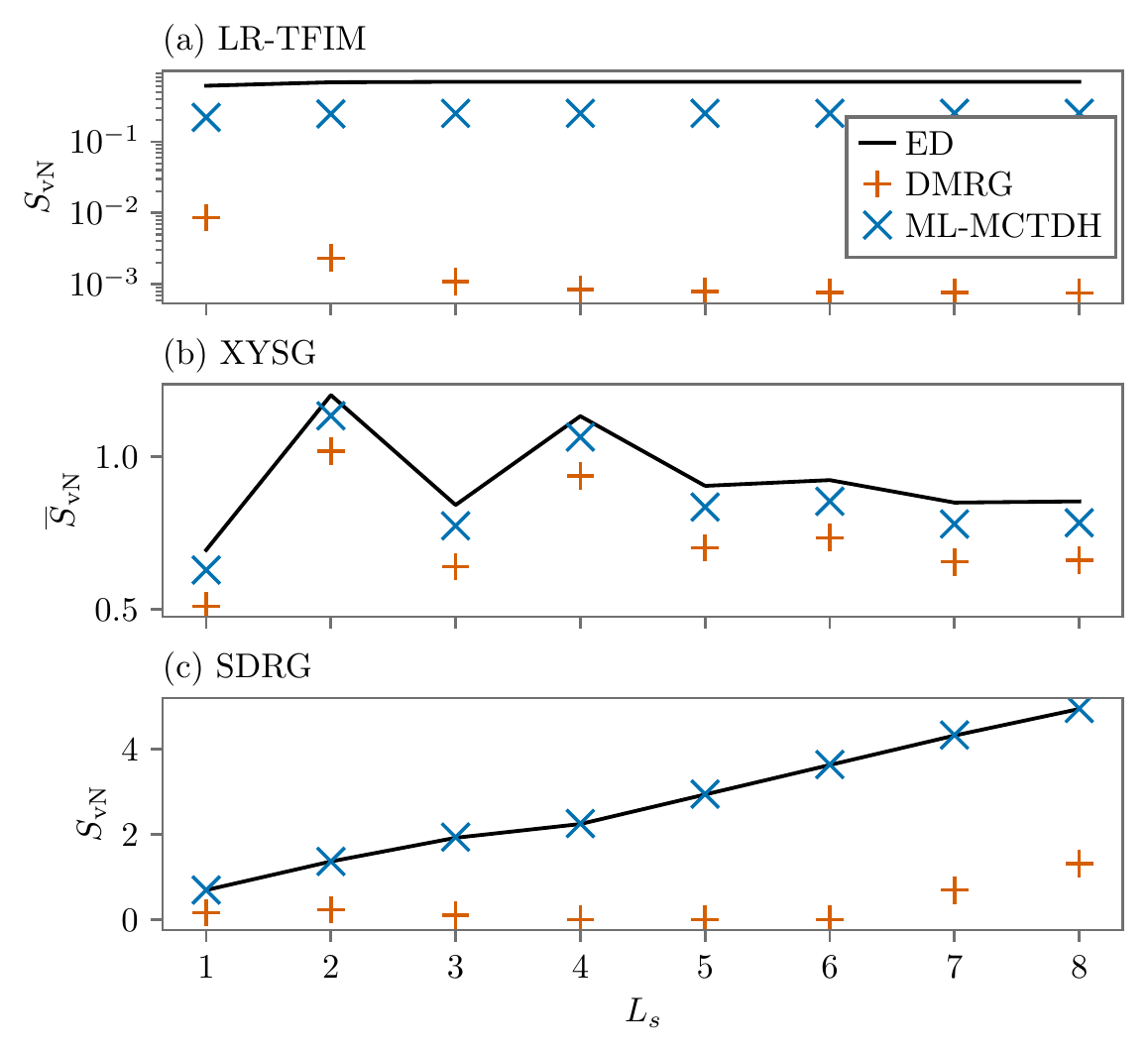}
    \caption{%
        VNEE~\eqref{eq:entropy} for the different models of $L=16$ spins in 1D as a function of subsystem size $L_s$.
        Panel (a) shows the result for the LR-TFIM with $h_x=0.5J$ and $h_z=0$, Panel (b) for the XYSG, averaged over $10$ disorder realizations, and panel (c) for the SDRG model.
    }
    \label{fig:entropy}
\end{figure}
One of the challenges when studying quantum many-body problems is the ability to capture nontrivial correlations.
Here, we use the connected correlation function~\cite{Parisi_1983_PRL50_1946_OrderParameterSpinGlasses,Verstraete_2004_PRL92_027901_EntanglementCorrelationsSpinSystems} which is defined as
\begin{equation}
    C_{\beta\beta}(i,j)=\braket{\sigma_{i}^{\beta}\sigma_{j}^{\beta}}-\braket{\sigma_{i}^{\beta}}\braket{\sigma_{j}^{\beta}},\quad\beta\in\lbrace x,y,z\rbrace\label{eq:connected-correlation-functions},
\end{equation}
to measure correlations in the system and characterize the magnetic ordering between spins $i$ and $j$.
Figure~\ref{fig:correlations} shows results for correlation functions defined in Eq.~\eqref{eq:connected-correlation-functions} for different spin models.
For the SR-TFIM in 1D and its 2D extension as well as the LR-TFIM we observe excellent agreement between all three methods as seen in panels (a)--(d).
The connected correlations in the $x$ direction denoted by $C_{xx}(i,j)$ were evaluated for disordered spin models by averaging over $10$ disorder realizations and over all unique spin pairings with $i<j$ corresponding to a given separation $r=|i-j|$.
The results are shown in panels (e)--(f).
In the case of XYSG, DMRG struggles to capture the correlations correctly as the low-energy spectrum exhibits many near-degeneracies, which are not well resolved by the DMRG algorithm such that it usually locks on to one of the first excited states.
Although this issue in DMRG can be mitigated by rescaling the Hamiltonian such that the energy splitting is increased this is not practical for larger systems.
ML-MCTDH, however, does not have any such issues.
When analyzing the correlations for the SDRG model [panel (f)] with respect to one of the center spins $c=\sfrac{L}{2}$, the methods agree with only a minor deviation for the value of $C(c,c+3)$ in the case of DMRG.
Due to the decay of the coupling constants towards the outer spins, the correlations will also quickly die off with increasing distance from the center spin.
Except for the minor deviation in the case of DMRG, all methods agree very well with each other.

In order to determine the entanglement of the ground state, we employ the von Neumann entanglement entropy (VNEE) $S_{\mathrm{vN}}$~\cite{Neumann_1927_NAWGMPK1927_273_ThermodynamikQuantenmechanischerGesamtheiten} of a subsystem $A$ with the remainder of the system,
\begin{equation}
    S_{\mathrm{vN}}=-\Tr\left[\rho_A\ln\left(\rho_A\right)\right]\label{eq:entropy}
\end{equation}
where $\rho_A$ is the reduced density matrix~\cite{Dirac_1930_MPC26_376_NoteExchangePhenomenaThomas} of the subsystem $A$.
Figure~\ref{fig:entropy} shows the VNEE defined in Eq.~\eqref{eq:entropy} for the spin models LR-TFIM in panel (a), XYSG in panel (b), and SDRG in panel (c) as a function of subsystem size $L_s$.
The subsystem $A$ was here chosen to consist of the $L_s$ left-most spins in the chain.
In the case of LR-TFIM, we chose the transverse field to be $h_x=0.5J$ such that the ground state is twofold degenerate due to the global spin-flip symmetry.
For a finite system, the ground state is expected to be a superposition state, which possesses non-negligible entanglement.
This behavior is correctly captured by ED and ML-MCTDH while DMRG yields a much lower entanglement as it converges to one of the degenerate states.
It is important to note that the exact superposition of both degenerate ground states is arbitrary in both ED and ML-MCTDH, which affects the absolute value of $S_{\mathrm{vN}}$ and explains the discrepancy between these two methods.
The disordered XYSG model is known to have area-law violation proportional to $S_{\mathrm{vN}}\propto\ln L_s$ which is not visible in Fig.~\ref{fig:entropy}(b) due to the small system size and low number of realizations.
Here, the discrepancy between the three methods can be attributed to the high amount of degeneracy in the low-energy spectrum.
The different algorithms lock on to different states and thus yield different results.
In the case of the SDRG model we observe a great agreement between ML-MCTDH and ED.
However, DMRG cannot describe the linear growth of entanglement $S_{\mathrm{vN}}\propto L_s$ due to the formation of distant singlet states, which then have to be entangled.

\section{Conclusions and Outlook}\label{sec:conclusions-and-outlook}

Solving a many-body problem with large system sizes requires sophisticated numerical methods that go beyond exact diagonalization.
Quantum Monte Carlo methods~\cite{Troyer_2005_PRL94_170201_ComputationalComplexityFundamentalLimitations} rely on the wave function spanning fewer relevant many-body configurations.
Other approaches represent the many-body wave function through an efficient compression of the state like with matrix product states, more general tensor networks or in some cases even neural networks~\cite{Melko_2019_NP15_887_RestrictedBoltzmannMachinesQuantum,Torlai_2019_PRL123_230504_IntegratingNeuralNetworksQuantum,Torlai_2020_ARCMP11_325_MachinelearningQuantumStatesNISQ}.
Despite the unquestionable success of these methods, they can fail for various reasons like the sign problem in quantum Monte Carlo methods~\cite{Troyer_2005_PRL94_170201_ComputationalComplexityFundamentalLimitations}, inefficiency of current quantum state compression in high-dimensional systems or due to the area-law violation.

In this paper, we propose an alternative computational method to explore many-body quantum spin models and specifically the case of disordered systems which are known to violate the area law in entanglement.
Focusing on the ground-state properties of prototypical many-body disordered spins models, ML-MCTDH achieves a remarkable accuracy in particular compared to conventional methods.
While MCTDH methods are regularly used to solve for complex wave packet dynamics problems, our paper is the first step in adapting these techniques to simulate a larger class of intricate many-body spin models.
One of the key advantages of using the multilayer version of MCTDH is its ability to treat large system sizes as well as degrees of freedom with many primitive basis states.
The latter aspect can be useful for simulating higher spin degrees of freedom such as $\text{SU}(n)$ physics~\cite{Manmana_2011_PRA84_043601_SuMagnetismChainsUltracold,Nataf_2014_PRL113_127204_ExactDiagonalizationHeisenbergSu} or higher spatial dimensions.
In future works, it will be interesting to  compare the performance of ML-MCTDH with existing numerical methods applied to higher dimensional spin lattices~\cite{White_1996_PRL77_3633_SpinGapsFrustratedHeisenberg,Evenbly_2011_JSP145_891_TensorNetworkStatesGeometry,Stoudenmire_2012_ARCMP3_111_StudyingTwoDimensionalSystemsDensity}.
MCTDH algorithms were originally built to study quantum dynamics.
Therefore, a natural next step would be to simulate many-body spin dynamics~\cite{Ng_2021_PRB103_134201_LocalizationDynamicsCentrallyCoupled} with these methods, which can be achieved very straightforwardly by switching to real time propagation.
We are convinced that ML-MCTDH can be a useful tool in this field of active research that includes intriguing topics like thermalization~\cite{Polkovnikov_2011_RMP83_863_ColloquiumNonequilibriumDynamicsClosed,Gogolin_2016_RPP79_056001_EquilibrationThermalisationEmergenceStatistical}, quench dynamics~\cite{Essler_2016_JSM2016_064002_QuenchDynamicsRelaxationIsolated,Guardado-Sanchez_2018_PRX8_021069_ProbingQuenchDynamicsAntiferromagnetic}, and optimal control~\cite{Khaneja_2005_JoMR172_296_OptimalControlCoupledSpin,Omran_2019_S365_570_GenerationManipulationSchrodingerCat}.
From a more technical point of view, there is also the scope for improving the scheme of building the different layers within ML-MCTDH:
One can potentially optimize this process by either using machine learning methods~\cite{Kaelbling_1996_JAIR4_237_ReinforcementLearningSurvey,LeCun_2015_N521_436_DeepLearning,Goodfellow_2016_DeepLearning,Sutton_2018_ReinforcementLearningSecondEdition,Melko_2019_NP15_887_RestrictedBoltzmannMachinesQuantum,Torlai_2019_PRL123_230504_IntegratingNeuralNetworksQuantum,Torlai_2020_ARCMP11_325_MachinelearningQuantumStatesNISQ} or spawning techniques~\cite{Mendive-Tapia_2020_JCP153_234114_RegularizingMCTDHEquationsMotion,Martinazzo_2021_AQ__CommentRegularizingMCTDHEquations,Martinazzo_2020_PRL124_150601_LocalinTimeErrorVariationalQuantum} and even combine them with tensor network methods~\cite{Larsson_2019_JCP151_204102_ComputingVibrationalEigenstatesTree}.
Thus, ML-MCTDH techniques can prove to be very a powerful alternative theoretical tool in modeling complex many-body (spin) systems.\\

\begin{acknowledgments}
    The authors acknowledge fruitful discussions with Henrik R.~Larsson and Frank Pollmann.
    This work is funded by the German Federal Ministry of Education and Research within the funding program ``quantum technologies - from basic research to market'' under Contract No. 13N16138.
    This work is supported by the Deutsche Forschungsgemeinschaft (DFG, German Research Foundation), SFB-925, project 170620586.
\end{acknowledgments}

\appendix
\section{Computation of Eigenstates}\label{app:computation_of_eigenstates}
Here, we discuss how ML-MCTDH can be applied to determine the many-body ground states of spin models by switching from real time to imaginary time propagation.
Solving the time-independent Schrödinger equation by diagonalization of the Hamiltonian matrix~\cite{Lin_1990_PRB42_6561_ExactDiagonalizationQuantumspinModels} is prohibitive for large systems.
Instead, eigenstates can be obtained by propagating an initial trial state according to the time-dependent Schrödinger equation in imaginary time $\tau=\imath t$~\cite{Kosloff_1986_CPL127_223_DirectRelaxationMethodCalculating}.
The evolution of the many-body wave function in the eigenbasis of the Hamiltonian reads $\ket{\Psi(\tau)}=\sum_n A_n(0) e^{-\tau E_n}\ket{\Psi_n}$.
After a sufficiently long propagation time the ground state becomes the dominant component of the instantaneous many-body wave function as long as its initial contribution $A_0(0)$ is not zero.
This scheme is feasible in the framework in (ML)-MCTDH as well~\cite{Beck_2000_PR324_1_MulticonfigurationTimedependentHartreeMCTDH} and has been applied for example to compute initial states in photodissociation studies~\cite{Manthe_1992_JCP97_3199_WavePacketDynamicsMulticonfiguration,Manthe_1992_JCP97_9062_MulticonfigurationalTimeDependentHartree,Manthe_1993_CPL211_7_WavepacketDynamicsFiveDimensions,Hammerich_1994_JCP101_5623_TimeDependentPhotodissociationMethyl}.
Since imaginary time propagation relies on the exponential damping of any contributions from excited states, often long propagation times are required in order to achieve adequate convergence towards the ground state.
The improved relaxation algorithm~\cite{Meyer_2003_TCA109_251_QuantumMolecularDynamicsPropagating,Wang_2014_JPCA118_9253_IterativeCalculationEnergyEigenstates,Wang_2015_JPCA119_7951_MultilayerMulticonfigurationTimeDependentHartree} employs a hybrid scheme consisting of imaginary time propagation and diagonalization to improve the convergence speed.
By applying the time-independent variational principle to the (ML)-MCTDH ansatz, one obtains an eigenvalue equation determining the top layer coefficients $A_{\ell_1,\ldots,\ell_d}^{(T)}(t)$ where $T$ is the number of layers.
The equations determining the SPFs on the lower layers could be solved iteratively, which, however, would result in highly nonlinear equations that are difficult to converge~\cite{Wang_2015_JPCA119_7951_MultilayerMulticonfigurationTimeDependentHartree}, similar to multiconfiguration consistent field theory~\cite{Hinze_1973_JCP59_6424_MCSCFMultiConfiguration}
Instead, the improved relaxation algorithm alternates between updating the top layer coefficients by solving the eigenvalue equation and imaginary time propagation to adapt the SPFs.
By always choosing the $n$th eigenvector to obtain a new set of top layer coefficients, the algorithm converges towards the $n$th eigenstate of the Hamiltonian.
Consequently, improved relaxation provides easy access to excited states, which would otherwise require to first compute and then project out lower lying states.
For the diagonalization involved we employ the implicitly restarted Lanczos method~\cite{Calvetti_1994_ETNA2_21_ImplicitlyRestartedLanczosMethod} via ARPACK~\cite{Lehoucq_1998_ARPACKUsersGuide}.

\section{DMRG Protocol}\label{app:dmrg_protocol}
In the present paper, we chose a sweep protocol of nine sweeps and allow a maximum bond dimension for the matrix product state of up to $1000$.
We followed a typical procedure of increasing the allowed maximum bond dimension with each sweep while always ensuring enough headroom between this value and the actual maximal bond dimension of the matrix product state.
For the first few sweeps we added a small noise term that improves the convergence and decrease its strength with each sweep.
Another important parameter is the cutoff that determines the actual bond dimension.
We followed best practice and started with a value of $10^{-6}$ at the beginning of the sweep protocol and decreased it rapidly with each sweep.
The last two sweeps were performed with a cutoff of $10^{-14}$, which ensures near exact accuracy.
We observe that even for a long chain of length $L=1024$ in the SR-TFIM, the final maximal bond dimension was only $14$, which is expected due to the short-range and homogeneous nature of the model.

\end{document}